\documentclass[10pt,preprint]{aastex}
\usepackage{natbib}
\usepackage{lscape}
\usepackage{ulem}
\usepackage[usenames]{color}
\slugcomment{Accepted for publication in ApJ Letters ~~\today}
\usepackage{epstopdf}

\newcommand       \msun        	{$M_{\odot}$}
\newcommand       \lsun      	{$L_{\odot}$} 
\newcommand       \hub {km~s$^{-1}$~Mpc$^{-1}$}

\newcommand       \mstar      {M$_{\star}$}

\newcommand	      \myr              {$M_{\odot}$~yr$^{-1}$}
\newcommand        \mic        	 {$\mu$m}

\newcommand \macs    {MACS1149-JD}
\newcommand \td      {$t_{exp}100$}
\newcommand \tc      {$t_{exp}300$}
\newcommand \mg      {$\widetilde{m_g}$}
\newcommand \md      {$\widetilde{m_d}$}
\newcommand \mge      {\widetilde{m_g}}
\newcommand \mde      {\widetilde{m_d}}

\begin{document}

\title{DUST FORMATION, EVOLUTION, AND OBSCURATION EFFECTS IN THE VERY HIGH-REDSHIFT UNIVERSE}

\author{Eli Dwek\altaffilmark{1},Johannes Staguhn\altaffilmark{1,2}, Richard G. Arendt\altaffilmark{1,3}, Attila Kovacks\altaffilmark{4}, Ting Su\altaffilmark{2}, \& Dominic J. Benford\altaffilmark{1}}
\affil{Observational Cosmology Lab., Code 665 \\ NASA Goddard Space Flight Center,
Greenbelt, MD 20771, \\ e-mail: eli.dwek@nasa.gov}
\altaffiltext{1}{Observational Cosmology Lab., Code 665, NASA at Goddard Space Flight Center, Greenbelt, MD 20771}
\altaffiltext{2}{Department of Physics and Astronomy, Johns Hopkins University, Baltimore, MD 21218}
\altaffiltext{3}{CRESST, University of Maryland Baltimore County, Baltimore, MD 21250}
\altaffiltext{4}{Astronomy Department, CalTech, Pasadena, CA 90025 and Astronomy Department, University of Minnesota, MN 12345}

\begin{abstract}
The evolution of dust at redshifts $z\gtrsim 9$, and consequently the dust properties, differs greatly from that in the local universe. In contrast to the local universe, core collapse supernovae (CCSNe) are the only source of thermally-condensed dust. Because of the low initial dust-to-gas mass ratio, grain destruction rates are low, so that CCSNe are net producers of interstellar dust. Galaxies with large initial gas mass or high mass infall rate will therefore have a more rapid net rate of dust production comported to galaxies with lower gas mass, even at the same star formation rate. The dust composition is dominated by silicates, which exhibit a strong rise in the UV opacity near the Lyman break. This ``silicate-UV break'' may be confused with the Lyman break, resulting in a misidentification of a galaxies' photometric redshift. In this paper we demonstrate these effects by analyzing the spectral energy distribution (SED) of \macs, a lensed galaxy at $z=9.6$. A potential 2mm counterpart of \macs\ has been identified with GISMO. While additional observations are required to corroborate this identification, we use this possible association to illustrate the physical processes and the observational effects of dust in the very high redshift universe.  
\end{abstract}

\section{INTRODUCTION}
Galaxies are the principle tracers of the global evolution of the star formation history, the metal enrichment, and the formation of dust in the universe. Their detection at very high redshifts offers a unique glimpse of these physical processes during the earliest epochs of galaxy evolution. Dust has been detected in several high-redshift galaxies \citep{gall11c}, the current record being held by the dusty quasar QSO J1120+0641 located at redshift $z = 7.12$~\citep{venemans12}. Dust attenuates the intrinsic spectrum of the stellar population, affecting the various physical quantities derived from UV-optical observations. Detection of dust through its extinction and emission is therefore essential in determining the properties of galaxies in the high-redshift universe. 

A  gravitationally magnified galaxy, designated MACS1149-JD was recently discovered at a photometric redshift of $z =9.6\pm$0.2 \citep{zheng12}. The massive lensing cluster (MACS J1149.6+2223), located at redshift 0.544, provides a magnification factor of $\mu=14.5\pm^{+4.2}_{-1.0}$. Assuming no extinction by dust, and correcting for magnification, the star formation rate (SFR), the total stellar mass, and luminosity derived from the observed UV-optical (UVO) flux are about $1.2\, \mu_{15}^{-1}$~\myr,  $1.5 \times10^{8}\, \mu_{15}^{-1}$~\msun, and  $2\times10^{10}\, \mu_{15}^{-1}$~\lsun, respectively, where $\mu_{15}$ is the lensing magnification normalized to the nominal value of 15 \citep{zheng12}. However, uncertainties in fitting the spectral energy distribution (SED) of this galaxy allow the possibility of a certain amount of extinction \citep{zheng12}. 

\begin{figure}
\includegraphics[width=3.2in]{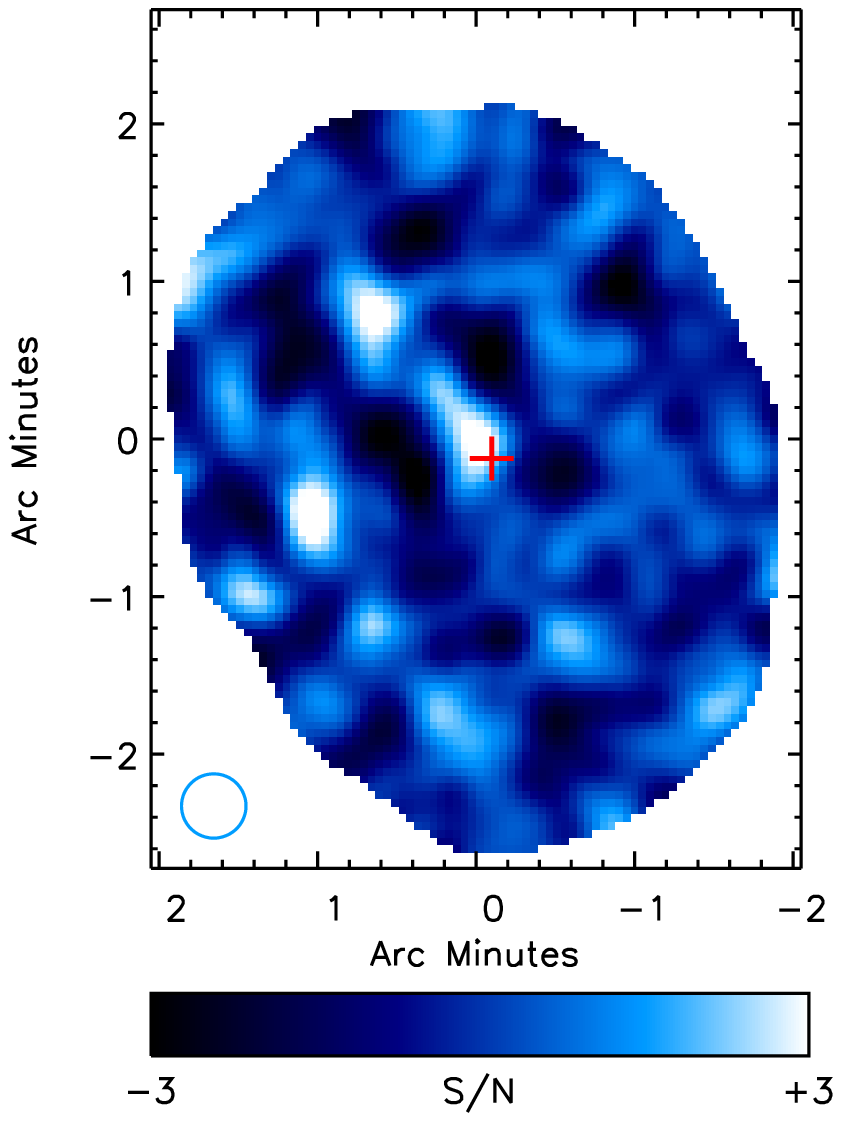}
\includegraphics[width=3.2in]{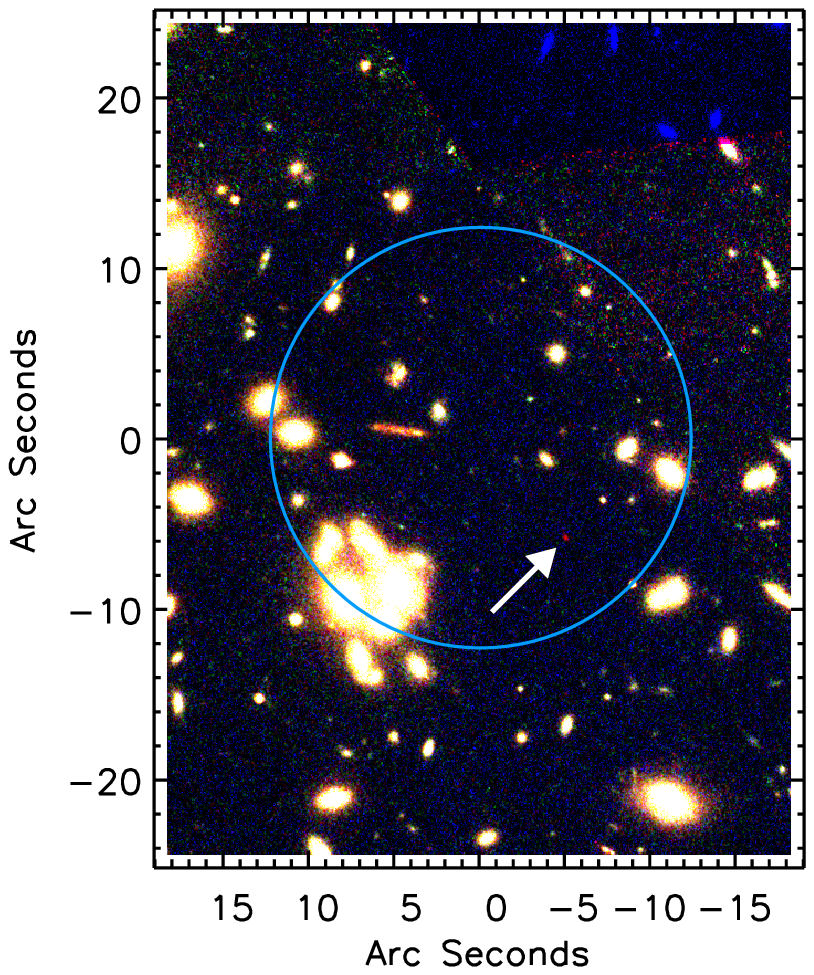}
\caption{\label{gismo}{\small  {\bf Left}: The GISMO 2 mm signal-to-noise map of the MACS1149-JD field. The
$4.1\sigma$ source  near the center of the field is
is $\sim8''$ (or $<0.5$ beam widths [FWHM] = $17.5''$) from the {\it Hubble}
location of MACS1149-JD (red cross), and therefore consistent with being the same source. 
Two other sources with S/N $>$ 3 are present in the field. 
 {\bf Right}: {\it HST} WFC3 image of MACS1149-JD at 0.85, 1.05, and 1.60 \mic\ (blue,
green, red respectively). MACS1149-JD (arrowed) can be seen as the only
extremely red source (a dropout at 0.85 and 1.05\mic). The field shown here
is much smaller than the GISMO field. The blue circle indicates the $24.7''$ FWHM
of the smoothed GISMO beam centered on the nominal location of the GISMO source.}}
\end{figure}

The Goddard-IRAM Superconducting 2~Millimeter Observer (GISMO) camera at the IRAM 30m telescope on Pico Veleta \citep{staguhn13}, Spain was used to probe whether MACS1149-JD has a rest frame far-infrared counterpart. The observations detected a  2~mm source with an intensity of $400\pm 98~\mu$Jy (see Figure~\ref{gismo}). The observed offset of $8''$ (less than half the GISMO beam) is entirely consistent with this source being at the {\it HST} detected position of MACS1149-JD. The rest wavelength is $\sim 190$~\mic, close to the peak of the dust emission in star forming galaxies. These observations strongly suggest that dust is present in this galaxy, however, further confirmations are needed to absolutely corroborate this association. Observations using the SHARC-2  submillimeter bolometer camera at the Caltech submillimeter observatory yielded a 350~\mic\ upper limit of $8.6\pm5.6$~mJy within an area consistent with the nominal position of \macs. 

For sake of the current analysis we will assume the validity of this association, adopt the observed 2~mm flux as originating from \macs, and demonstrate the unique properties of any dust that may be forming in the high redshift universe.
All fluxes and results of this paper are presented in the galaxy's rest frame with the adopted cosmological parameters $H_0 = 67$~\hub, $\Omega_M=0.315$, and $\Omega_{\Lambda} = 0.685$ \citep{planck-ade13}, and a lensing magnification factor of 15. With these cosmological parameters the age of the universe is $\sim 500$~Myr at $z=9.6$.

\section{MODELING THE SED OF MACS1149-JD} 
In the absence of any starlight,  dust would be in thermal equilibrium with the cosmic microwave background (CMB). This effect can be important at high redshifts where starlight and the CMB can have comparable energy densities \citep{da-cunha13}.
The GISMO flux is the differential 2~mm flux with respect to the CMB emission. Therefore, the observed 2~mm flux consists of only reradiated starlight. 
 We characterize the IR spectrum by $T_d$, the temperature that would be attained by the dust in the absence of the CMB, and by a mass absorption coefficient $\kappa(\lambda)$, that has a $\lambda^{-\beta}$ ($\beta=1.5$) dependence at far-IR wavelengths and a value of 1.5~cm$^2$~g$^{-1}$ at $\lambda = 850$~\mic\ \citep{kovacs10}. The specific IR luminosity in the galaxy's rest frame is then given by:
\begin{equation}
\label{ }
L_{\nu}(\lambda)  =  4 M_d\, \kappa(\lambda)\, \left[\pi B_{\nu}(\lambda, T_{eff})-\pi B_{\nu}(\lambda, T_{cmb})\right] 
\end{equation}
where $T_{eff}  =  (T_d^{4+\beta} + T_{cmb}^{4+\beta})^{1/4+\beta}$, $M_d$ is the dust mass, and $T_{cmb} = (1+z)\, 2.73 \approx 29$~K is the CMB temperature at $z=9.6$. 
$L_{\nu}(\lambda)$ is a function of the dust temperature and mass. With $L_{\nu}(\lambda=190$\mic) normalized to the observed GISMO flux, the dust temperature $T_d$ and dust mass are inversely related. 

\begin{figure*}
\begin{center}
\includegraphics[width=3.2in]{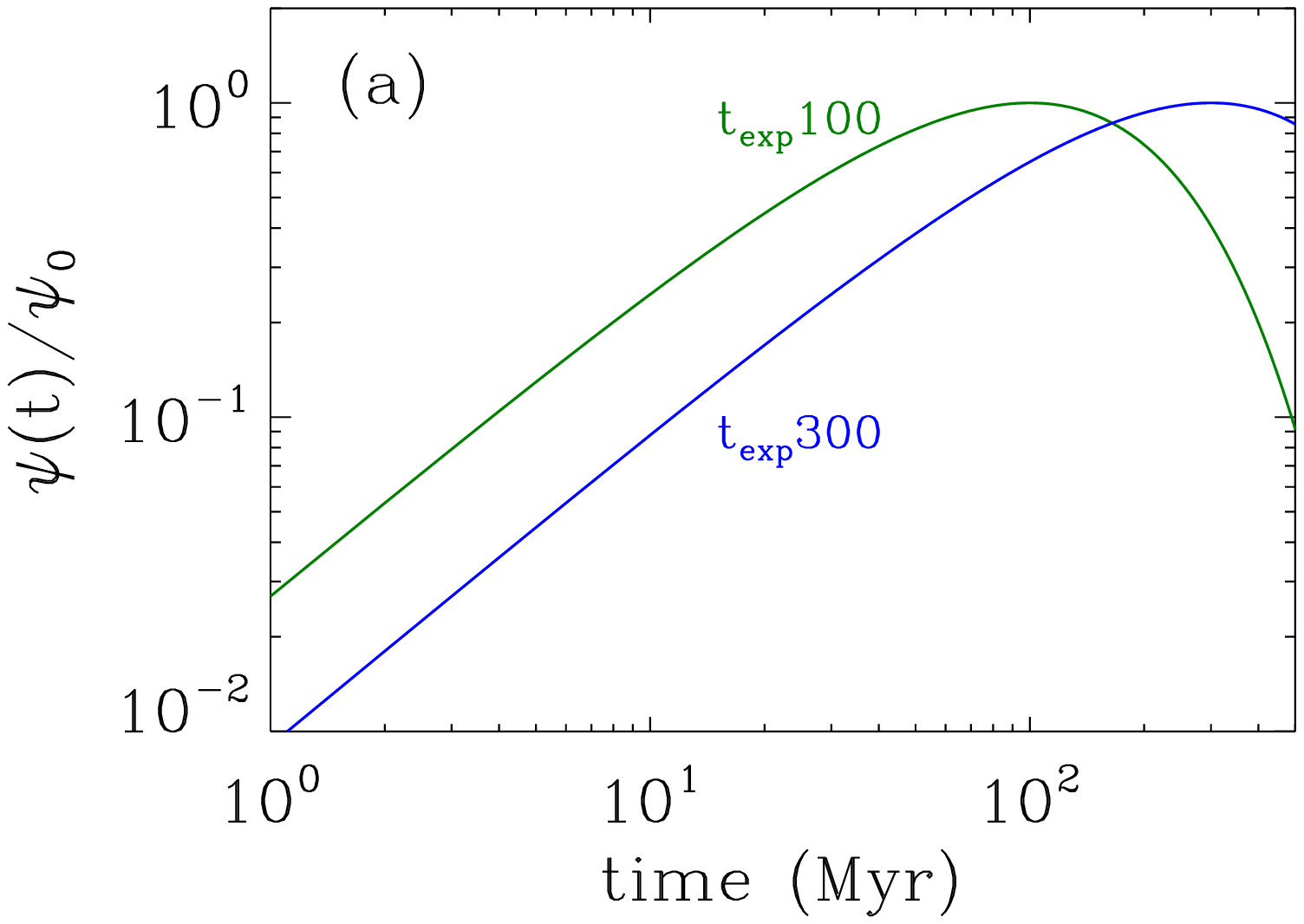}
\includegraphics[width=3.2in]{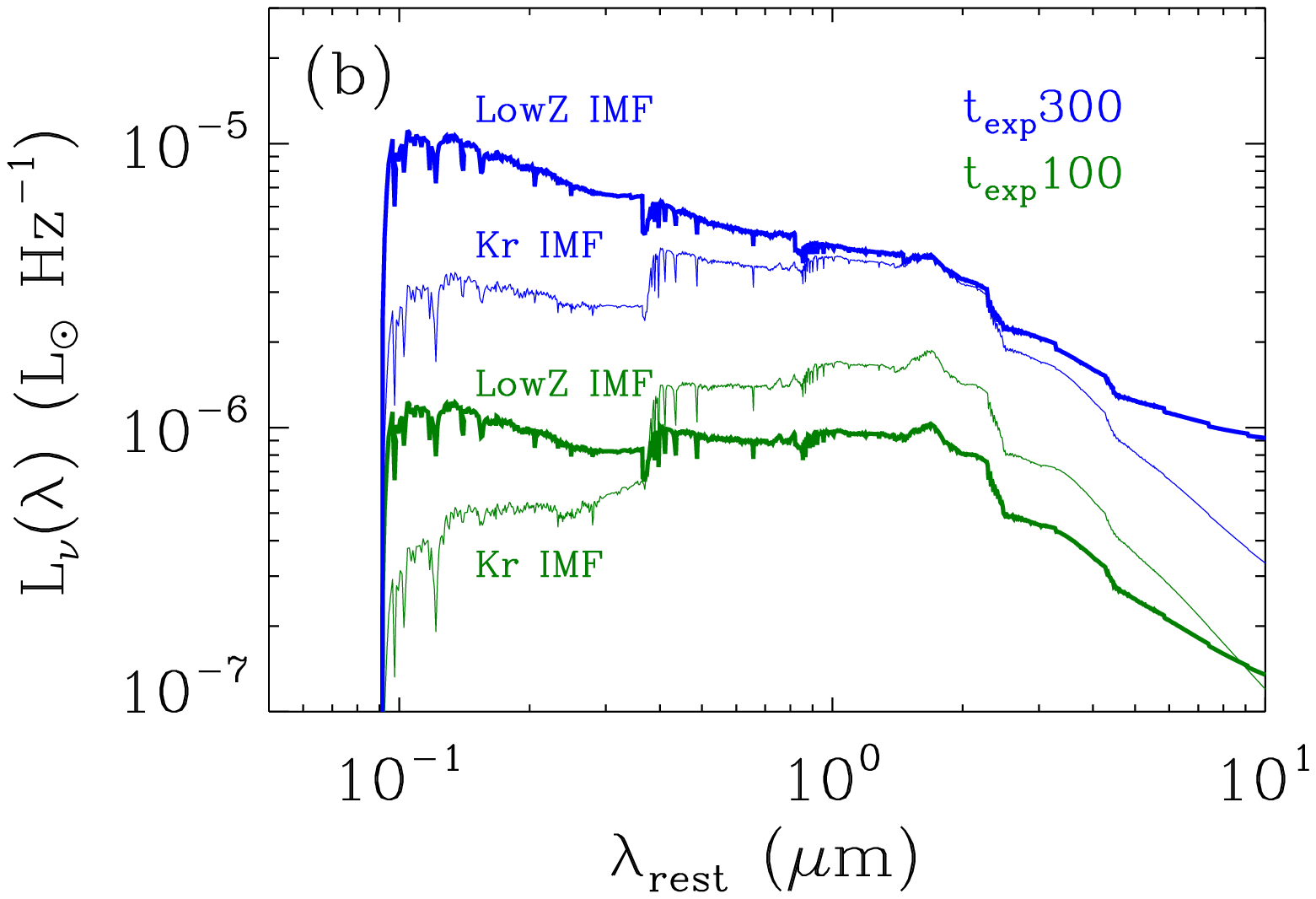}\\
~~~ \\
\includegraphics[width=3.2in]{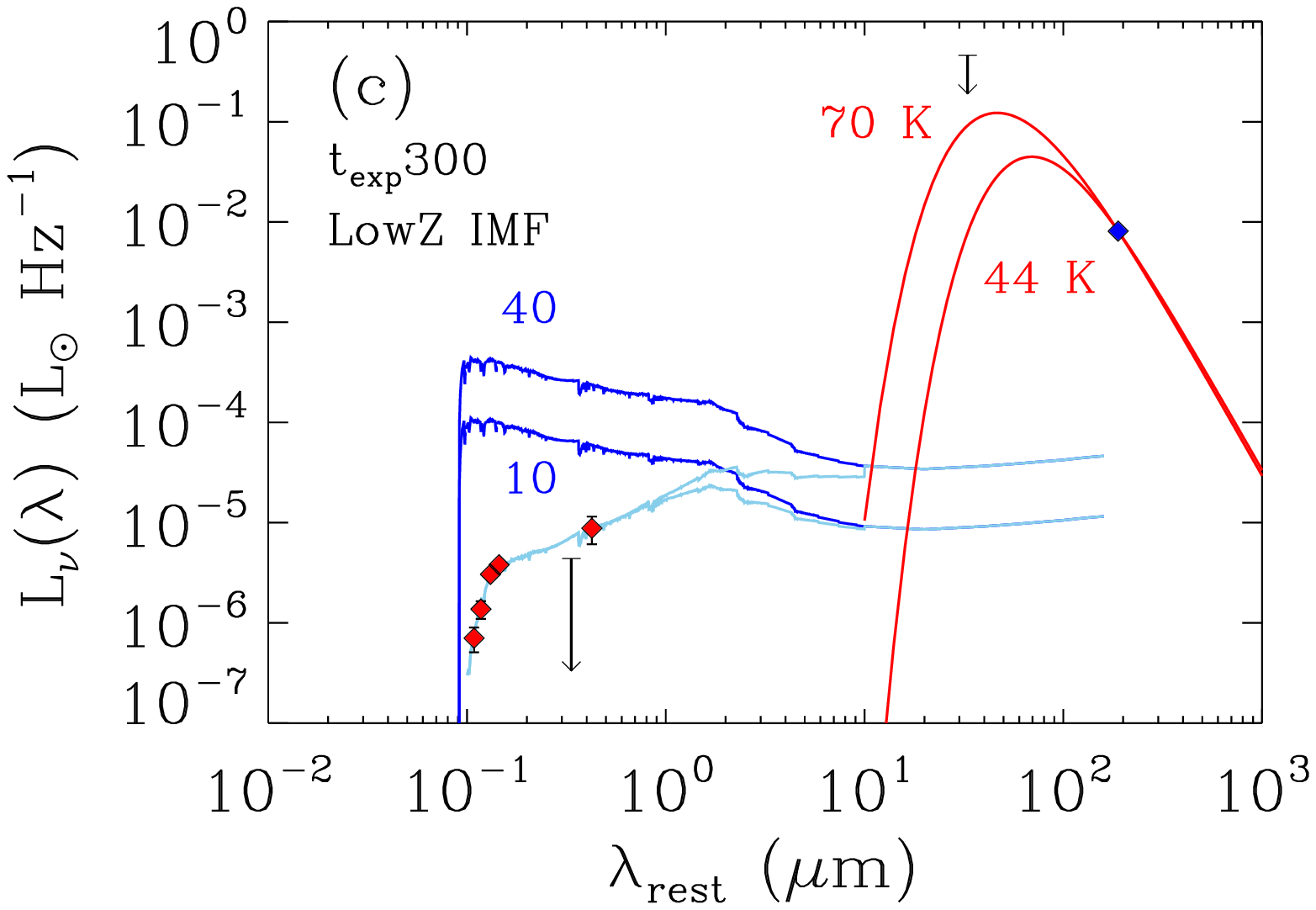}
\includegraphics[width=3.2in]{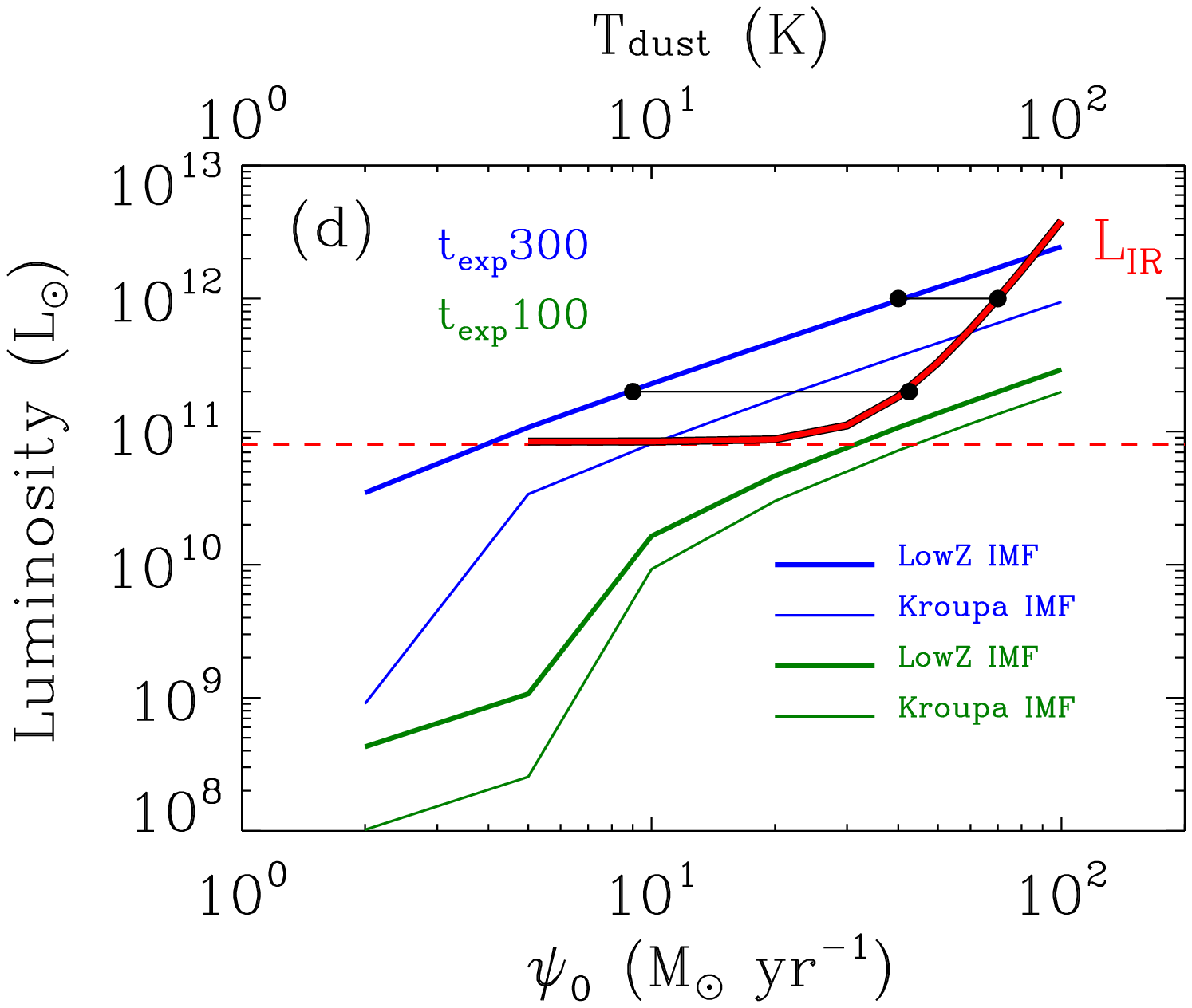}\\
\includegraphics[width=3.2in]{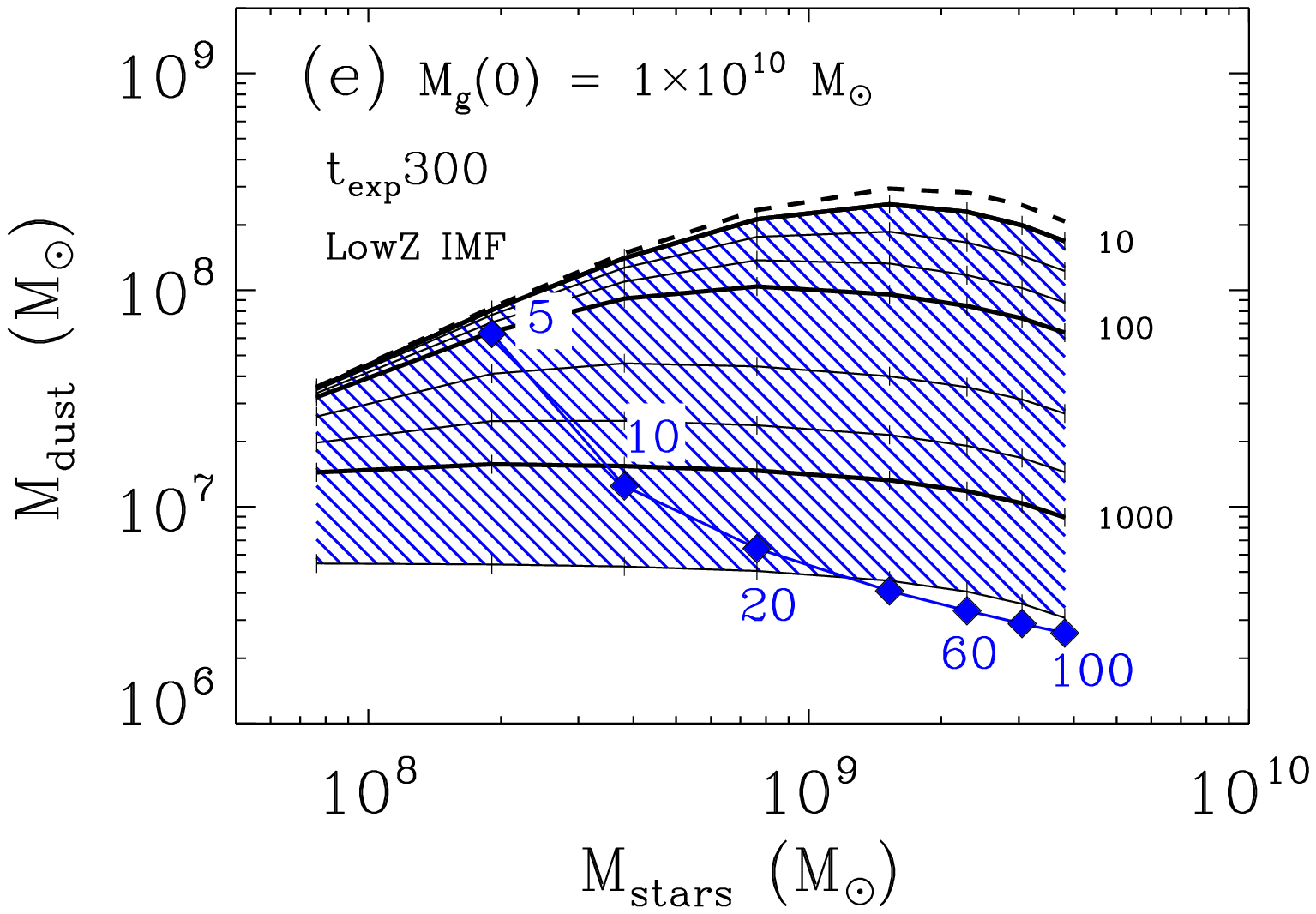}
\includegraphics[width=3.2in]{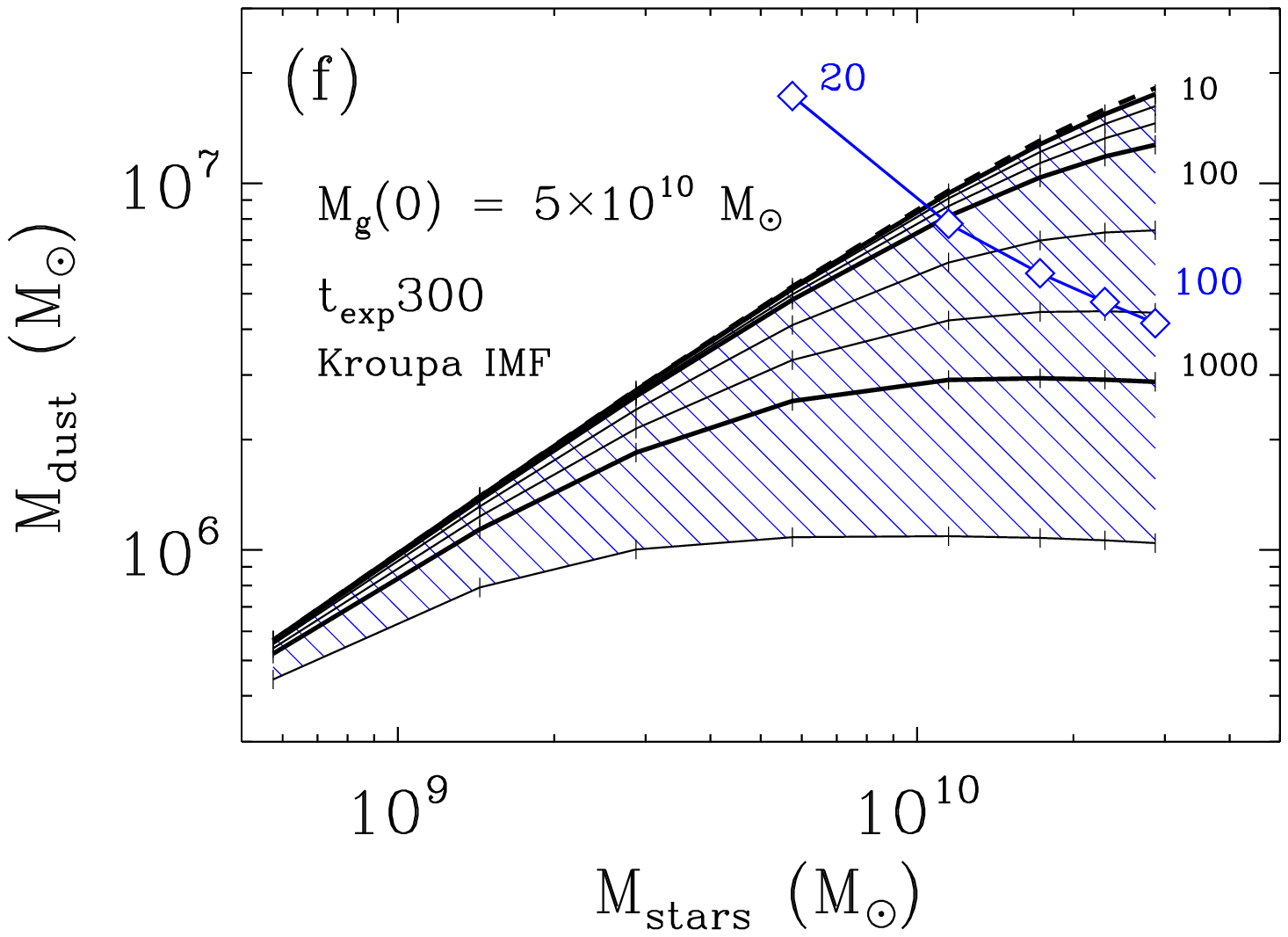}\\
\caption{\label{SEDfit}{\small(a) The star formation rates as a function of cosmic age considered in this study have time scales of $t_{exp} = 100$ or 300~Myr. At the time when \macs\ is observed ($t = 500$~Myr) the SFR has fallen to  0.092 or 0.86 of the peak SFR, $\psi_0$; (b) the spectrum of the galaxy at $t=500$~Myr for the different SFRs shown in (a) for the Low-metallicity and Kroupa IMF. 
 (c) models for the intrinsic spectrum must satisfy the energy constraint, requiring the absorbed stellar luminosity to be equal to the reradiated IR Luminosity. The two examples depicted in the figure are characterized by values of $\psi_0 = 10$ and 40~\myr, and  corresponding dust temperatures of $T_d=44$ and 70~K; (d) depiction of SF scenarios that satisfy the energy constraint. The scenarios are characterized by their value of $t_{exp}$, $\psi_0$, and the stellar IMF. The IR luminosity (red curve) is characterized by its dust temperature. The dashed red line indicates the lower limit on $\psi_0$ for the different SF scenarios (see Table~1). The filled circles on the two horizontal lines illustrate the connection between $\psi_0$ and $T_d$ for the two cases shown in Figure~2c; (e)-(f) Comparison of modeled stellar and dust masses (hatched regions) and SED determined masses (diamonds) as a function of $\psi_0$ (vertical ticks) and \mg\ (black lines). See text for details.}} 
\end{center}
\end{figure*}

The presence of dust significantly affects the properties of \macs\ derived under the assumption that it is dust free, and constrains viable star formation scenarios for this galaxy. The panels of Figure~\ref{SEDfit} depict the analysis used to constrain the star formation history (SFH), the amount of dust obscuration and resulting infrared emission, and the mass of radiating dust, based on the available UVO data and adopted far-IR flux.  
Two star formation histories, characterized by a delayed exponential star formation rate (SFR), $\psi(t) =\psi_0\, (t/t_{exp})\, \exp(-t/t_{exp}+1)$, were used in the analysis with values of $t_{exp}=$100 and 300~Myr (hereafter models \td\ and \tc, respectively, see Figure~\ref{SEDfit}a). 

Intrinsic stellar spectra were calculated using the stellar population synthesis code P\'EGASE \citep{fioc97} with low-metallicity (LowZ) and Kroupa (Kr) stellar initial mass functions (IMFs) \citep{marks12,kroupa01}. The LowZ IMF is more heavily weighted towards the formation of massive stars, and its stellar population produces more UV radiation compared to its Kroupa counterpart. Its choice is motivated by the theoretical argument that inefficient cooling inhibits the fragmentation of protostellar clouds into smaller units \citep{bromm04}, and by observed variations of the IMF with stellar metallicity \citep{marks12,kroupa13}. Figure ~\ref{SEDfit}b shows the spectra for $\psi_0=1$~\myr. 

For each star formation (SF)  scenario, characterized by $t_{exp}$, $\psi_0$, and a stellar IMF, we calculated the amount of dust absorption needed to reproduce the observed UVO fluxes. The ratio between the observed and intrinsic flux is the escape probability at that given wavelength, which for a homogeneous sphere is given by \citep{lucy91,cox69,osterbrock06}:
\begin{eqnarray}
P_{esc}(\tau, \omega) & = & P_{esc}(\tau,0)\left[1-\omega + \omega\, P_{esc}(\tau,0)\right]  \\ \nonumber 
P_{esc}(\tau,0) & = & {3\over 4\tau}\, [1-{1\over 2\tau^2} + ({1\over \tau} + {1\over 2\tau^2})\, \exp(-2\tau)] 
\end{eqnarray}
where $\tau(\lambda)=(3/4)M_d\, \kappa(\lambda) /\pi R^2$ is the optical depth, and $\omega(\lambda)$ is the dust albedo.
The second expression corresponds to the case when $\omega=0$. The albedo of small dust grains ($a \lesssim 0.01$~\mic) is $\lesssim 0.2$ and $P_{esc}(\tau,0) \approx 0.1$ for most scenarios. With these parameters, scattering will increase the escape probability by at most 20\%, having only a minor effect on the conclusions of this paper. 

We assumed that the IR emission is only powered by stellar radiation, so that the dust giving rise to the IR emission is also responsible for the attenuation of the stellar emission. 
The reradiated IR luminosity, given by the integral of $L_{\nu}(\lambda)$ in eq. (1),  must then be equal to that the starlight energy absorbed by the dust:
\begin{equation}
\label{ }
L_{abs} = \int\ L_{\nu}^{int}(\lambda)\, [1-P_{esc}(\lambda)]\, d\nu
\end{equation}
where $L_{\nu}^{int}$ is the intrinsic stellar luminosity. 
Figure~\ref{SEDfit}c illustrates this principle of energy balance for model \tc --LowZ for values of $\psi_0=20$ and 40~\myr. The dust must correspondingly radiate at temperatures of 44 and 70~K, with dust masses of $1.2\times10^7$ and $4.0\times10^6$~\msun, respectively.

Not all SF scenarios can fulfill the energy constraint. Figure~\ref{SEDfit}d plots the  IR luminosity as a function of $T_d$, and the absorbed stellar luminosity as a function of \ $\psi_0$. Energy conservation requires $L_{IR}$ (red curve) to be equal to $L_{abs}$ (green and blue curves). The GISMO detection sets a lower limit on the dust luminosity regardless of the dust temperature. This constraint sets a corresponding lower limit on $\psi_0$ for each SF scenario (see Table 1 for details). 

An additional constraint on viable SF scenarios is that the dust mass giving rise to the UVO attenuation and the IR emission be produced within the allotted time of $< 500$~Myr, taking possible grain destruction processes into account \citep{dwek07b}. 

\section{THE EVOLUTION OF DUST AT HIGH REDSHIFTS} 
Dust in galaxies is produced in the ejecta of explosive core-collapse supernovae (CCSNe) and in the quiescent winds of asymptotic giant branch (AGB) stars which are not massive enough to end their life as CCSNe. Refractory elements locked up in dust are returned to the gas phase by thermal and kinetic sputtering and evaporative grain-grain collisional processes in supernova blast waves \citep{jones04}. Star formation removes both, gas and dust from the ISM. The equations governing the evolution of the dust and gas masses in the ISM of galaxies were presented in \cite{dwek11a}. 
At high redshifts, CCSNe dominate the rate of dust production at a rate given by $dM_d/dt = \widetilde{Y_d}\, R_{SN}$, where $\widetilde{Y_d}$ is the IMF-averaged dust yield in CCSNe \citep[see][Table~2 for the maximum attainable dust yield as a function of stellar mass]{dwek07b}, and $R_{SN} \propto \psi_0$ is the IMF-dependent rate of CCSNe. CCSNe are also the main source of grain destruction in the ISM, at a rate given by $M_d/\tau_d$, where $\tau_d(t)$ is the dust lifetime, given by \citep{dwek80}:
\begin{equation}
\label{ }
 \tau_d = {M_d(t) \over \mde(t)\, R_{SN}(t)} = {M_g(t) \over \mge\, R_{SN}(t)}
\end{equation}
where \md$(t)$ is the time-dependent total mass of refractory elements locked in dust that is returned back to the gas phase by a single supernova remnant, \mg $ \equiv$ \md$(t)/Z_d(t)$ is the effective mass of interstellar gas that is totally cleared of dust, where $Z_d(t)\equiv M_d/M_g$ is the dust-to-gas mass ratio in the ISM, and $M_g$ is the total gas mass. The dust destruction rate is then given by:
\begin{equation}
\label{ }
{dM_d\over dt} = \mde(t)\, R_{SN}(t) = Z_d(t) \mge\, R_{SN}(t)
\end{equation}
 
The value of \mg\ is approximately time-independent and about equal to $10^3$~\msun\ \citep{dwek07b}, and $Z_d \approx 0.007$ in the solar neighborhood \citep{zubko04}. The total mass of dust destroyed by a single SNR is therefore $\sim 7$~\msun, significantly larger than the total $\sim 1$~\msun\ of condensible elements in their ejecta. In the local universe, CCSN are therefore net destroyers of interstellar dust.  

In a dust-free galaxy, CCSNe are obviously net sources of interstellar dust. There is therefore a limiting value of $Z_d \approx 1\times 10^{-4}$ for which the grain destruction rate by SNR is balanced by their formation rate, assuming a 10\% condensation efficiency in the ejecta. The initial rate of dust enrichment in a galaxy depends therefore on: (1) the stellar IMF - a top-heavy IMF will have a higher IMF-averaged dust yield; and (2) the evolution of the gas mass in the galaxy, which in the instantaneous recycling approximation is given by:
\begin{equation}
\label{ }
M_g(t) = M_g(0) - \left[{\left<m_{ej}\right>\over \left<m\right>}\right]\, \int_0^t\psi(t')\, dt' + \left({dM_g\over dt}\right)_{inf}
\end{equation}
where $M_g(0)$ is the initial gas mass, $\left<m_{ej}\right>$ and $\left<m\right>$ are, respectively, the IMF-averaged ejecta and initial stellar mass, and $(dM_g/dt)_{inf}$ is the infall rate of the gas into the galaxy.

Dust masses derived from evolutionary models depend on $\psi_0$ and \mg, whereas dust masses derived from fitting the UVO-IR SED depend only on $\psi_0$. Figures~\ref{SEDfit}e-f depict these two distinctly derived dust masses as a function of the stellar mass derived from the corresponding population synthesis and dust evolution models at $t=500$~Myr, the epoch of observations. Figure~\ref{SEDfit}e represents the results for the LowZ IMF and an initial gas mass of $1\times 10^{10}$~\msun, and Figure~\ref{SEDfit}f for the Kroupa IMF and an initial gas mass of $5\times 10^{10}$~\msun\ with no infall. The solid black lines show the relation between $M_d$ and stellar mass for the dust evolutionary models with selected lines labeled by the value of \mg\ (in \msun).  Tick marks along the lines indicate the values of $\psi_0$. The models were calculated assuming that all refractory elements condensed in the mass outflow from AGB stars, but that only 10\% did in the ejecta of CCSN.  
The thick dashed lines show the derived dust masses in the absence of any grain destruction. The rate of grain destruction depends on the dust-to-gas mass ratio in the ISM, and therefore on the choice of the initial mass of the gas reservoir. The line connecting the filled and open diamonds represents the relation between the dust mass derived from the SED fitting to the stellar mass derived from the population synthesis models. The diamonds are labeled by the value of $\psi_0$. The hatched region gives the dust masses for all combinations of $\psi_0$ and \mg\ considered in this paper. Viable SF scenarios are those for which dust masses derived from SED fitting are equal to those derived from the evolutionary models (hatched regions). The resulting constraints on the SF scenarios and dust and stellar masses are summarized in the 3rd and 4th column of Table~1.

In general, SF scenarios characterized by a LowZ IMF produce the inferred dust mass at lower SFR and lower stellar masses. 
Scenarios that require low values of \mg\ imply dust lifetimes that are significantly larger than those expected for a homogeneous ISM, implying a clumpy ISM \citep{dwek79,dwek07b}. 

\begin{figure}
\begin{center}
\includegraphics[width=5.0in]{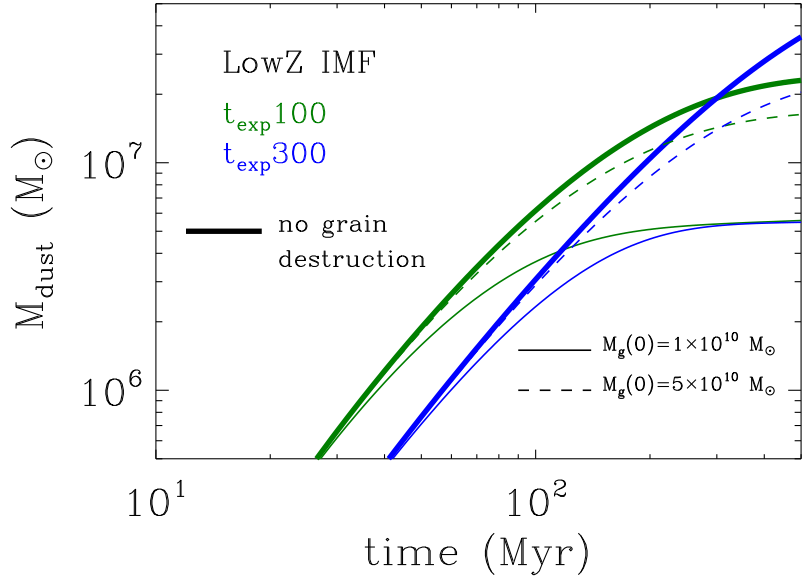}
\caption{\label{dustvol} \small{The evolution of the dust mass for a closed box model with a low-metallicity IMF  and $\psi_0=2$~\myr. The bold solid lines assume no grain destruction. The thin solid and dashed lines are models with different initial gas mass. The evolution of the gas mass determines the rate of grain destruction. Large gas masses imply low dust-to-gas mass ratios and lower rate of grain destruction [eqs. (5) and (6)]. The figure also shows that AGB stars do not contribute to the dust mass.}}
\end{center}
\end{figure}

Figure~\ref{dustvol} depicts the evolution of the dust for different values of the SFR and the stellar IMF, highlighting the role of the gas mass in determining the dust production rate. The models assume no infall, so the gas mass is only determined by its initial value and net consumption rate by stars. Models with large initial gas masses produce dust at a higher rate than those with a  lower initial gas mass. Eventually, the dust-to-gas mass ratio will reach the critical value of $\sim 0.001$ after which the rate of grain destruction by CCSN exceeds the rate of dust formation in their ejecta. Galaxies with large initial gas mass have therefore a faster rate of dust enrichment. However, they will also be rarer objects since lower mass galaxies are more readily assembled at $z\approx 10$ than the more massive ones \citep{somerville08, behroozi13}. 

\section{THE ULTRAVIOLET SILICATE BREAK}
\begin{figure}
\begin{center}
\includegraphics[width=5.0in]{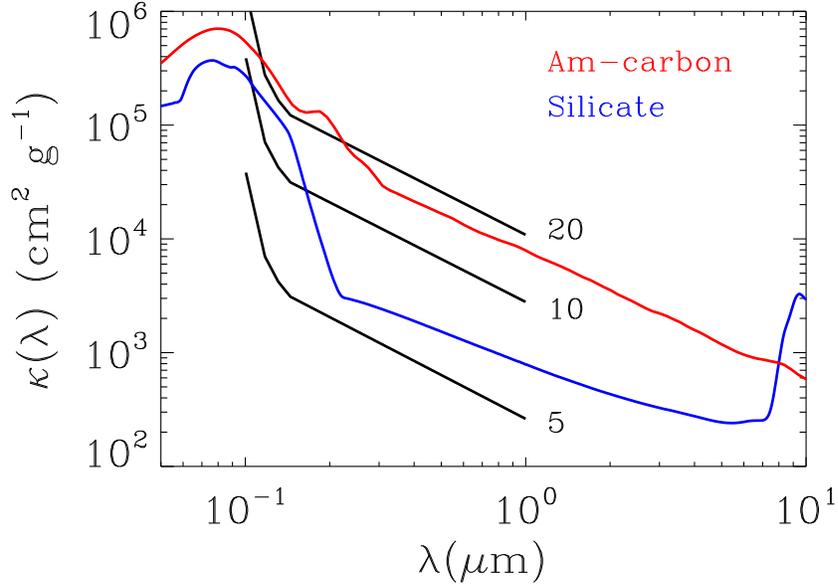}
\caption{\label{kappa} \small{The wavelength dependence of $\kappa(\lambda)$, derived from forcing the intrinsic spectra to reproduce the observed galaxy's UVO spectrum,  is compared to the mass absorption coefficient of 0.01~\mic\ radius astronomical silicate and amorphous carbon grains. The curves are labeled by the value of $\psi_0$ in units of \myr. For a given optical depth and dust mass, $\kappa$ is proportional to the projected area of the disk. Plotted values assume an area of 1~kpc$^2$.
The derived mass absorption coefficient exhibits a similar rise at UV wavelengths as the small silicates grains. Young galaxies with a silicate dominated population of interstellar dust grains will therefore have a break in their spectrum around 0.15~\mic, which can be confused with the Lyman break.}}
\end{center}
\end{figure}

The galaxy's intrinsic spectrum was forced to reproduce the observed UVO emission (see Figure~\ref{SEDfit}c). It is therefore of interest to examine how the derived mass absorption coefficient of the dust compares with that of astrophysical dust grains. Figure~\ref{kappa} compares the wavelength dependence of $\kappa(\lambda)$ with that of 0.01~\mic\ radii silicate and amorphous carbon grains, with optical constants from \cite{li01}. 
The mass absorption coefficient shows a sharp rise at $\lambda \lesssim 0.15$~\mic. The rise is similar to that of the silicate grains but at shorter wavelengths. A good match would require the galaxy to be at a somewhat lower redshift than that inferred from the UVO SED fitting. This suggests that young galaxies, in which silicates are the dominant dust component, will exhibit a sharp drop in their spectrum around 0.15~\mic\ caused by this rising silicate UV opacity. This ``UV silicate break'' could be mistaken for an intrinsic Lyman break at $\lambda = 0.0912$~\mic, leading to an overestimate in the photometric redshift determination of galaxies. Such galaxies may also exhibit a break in their IR spectrum that is associated with the 9.7 and 18~\mic\ silicate absorption features. This ``IR silicate break'' can be used to photometrically select galaxies in the $z\approx 1-2$ range \citep{takagi05}, or to estimate the photometric redshift of high-$z$ galaxies.

\section{SUMMARY}
The evolution of dust at redshifts above $\sim 10$ differs fundamentally  from that at lower redshifts: (1) at these redshifts, CCSNe are the dominant sources of interstellar dust. The enrichment of the ISM with carbon and carbon dust is delayed until the most efficient carbon producing AGB stars have evolved off the main sequence. The composition of the ISM and its dust may therefore be significantly different from that in lower redshift galaxies \citep{dwek98}. The redshift interval over which this effect can be observable may be quite narrow, depending on the stellar IMF and the mass range of carbon rich stars; (2) at low metallicities, hence low dust-to-gas mass ratios, CCSNe are net producers of interstellar dust. This is in sharp contrast to lower redshift galaxies, where CCSNe destroy more dust than they produce, which has led to the suggestions that most of the interstellar dust in galaxies must be reconstituted by cold accretion in the ISM \citep{dwek80, zhukovska08a, draine09, calura10, valiante11, gall11a, zhukovska14}; (3) the efficient production of dust in the high-z universe is also greatly facilitated  by the fact that at low-metallicities the stellar IMF is biased towards the formation of massive stars, further reducing the relative importance of lower mass AGB stars to the production of carbon in the early universe. 

Analysis of the SED of \macs, including the 2~mm (190~\mic\ restframe) flux from its GISMO counterpart, shows that dust formation proceeds at rapid rate at the earliest stages of galaxy evolution. Dust masses of $\sim (2-7)\times 10^7$~\msun\ can be readily produced within 500~Myr of evolution with SFR as low as $5-8$~\myr, without the need to resort to growth in the ISM.  This dust exhibits a sharp rise in opacity at UV wavelengths around 0.15~\mic, characteristic of very small silicate grains expected to form in the ejecta of CCSNe \citep{nozawa10,cherchneff10}.
 The sharp rise produces a "UV silicate break" in the spectra of very young galaxies, which can be mistaken for a Lyman break, leading to an overestimate of their photometric-based redshifts.  
The existence of high-redshift, actively star-forming, galaxies with little dust, such as Himiko at $z = 6.62$ \citep{ouchi13}, shows that the dust abundance in galaxies strongly depends on their star formation history, the efficiency of grain destruction processes, and the presence of gas infalls and dusty outflows. Our analysis shows that dust can form following the death of the first massive stars, but its survival and abundance will depend on the particular physical processes operating in each individual galaxy. 
Finally, independent confirmation of the association of the GISMO~2mm source with \macs, will render this galaxy the youngest dust forming galaxy, with profound implications for the onset of radiative, thermal, and chemical processes in the early universe.

\newpage
\clearpage
\begin{deluxetable}{lllll}
\tabletypesize{\small}
\tablecaption{Viable star formation scenarios following successive model constraints \tablenotemark{1,2}}
\tablehead{
\colhead{ } &
 \colhead{ } &
  \colhead{Energy constraint} & 
   \multicolumn{2}{c}{Mass production constraint: $M_d(SED fit) = M_d(evol)$ } \\
     \cline{4-5} \\   
\colhead {SF scenario} &
 \colhead{ } &
 \colhead{$L_{abs} = L_{IR}$} &
     \colhead{$M_g(0)=1\times10^{10}$~\msun}  & 
     \colhead{$M_g(0)=5\times10^{10}$~\msun} 
  }
  \startdata 
     $t_{exp}=100$~Myr & Low-Z IMF  & $\psi_0 \gtrsim 30$ & $\psi_0 \gtrsim 40$                & $\psi_0 \approx 40-50$                \\
                       &            &                     & \mstar$ \gtrsim 7\times10^8$   &  \mstar$\approx (7-10)\times10^8$  \\
                       &            &                     & $M_d \lesssim 6\times10^7$       & $M_d\approx (7-3)\times10^7$       \\
                       & Kroupa IMF & $\psi_0 \gtrsim 45$ & no viable models                   &  no viable models                  \\
 & & & & \\
\hline
 & & & & \\
     $t_{exp}=300$~Myr & Low-Z IMF  & $\psi_0 \gtrsim 4$  & $\psi_0 \approx 5-35$              &   $\psi_0 \approx 5-8$              \\
                       &            &                     & \mstar$\approx (2-12)\times10^8$   & \mstar$\approx (2-3)\times10^8$      \\ 
                      &            &                      & $M_d\approx (70-5)\times10^6$      & $M_d\approx (7-2)\times10^7$         \\      
                       & Kroupa IMF & $\psi_0 \gtrsim 10$ &  no viable models                  & $\psi_0 \gtrsim 35$                         \\
                       &            &                     &                                    & \mstar$ \gtrsim 1\times10^{10}$             \\ 
                       &            &                     &                                    & $M_d \lesssim 9\times10^6$          \\ 
\enddata
\label{tab:models}
\tablenotetext{1}{Values of $\psi_0$ are in units of \myr.}
\tablenotetext{2}{Values of the stellar and dust masses, \mstar\ and $M_d$, respectively, and are in units of \msun.}
\end{deluxetable}

{\bf Acknowledgements} - 
This work was supported through NSF ATI grants 1020981 and 1106284 (J.S., T.S., A.K. and the GISMO observations). IRAM is supported by INSU/CNRS (France), MPG (Germany) and IGN (Spain). ED and R.G.A. acknowledges support of NASA-ROSES-ATP2012. We acknowledge the comments made by the referee which have led to a more detailed discussion on the origin of dust in the early universe. E.D. thanks Rachel Somerville for a helpful discussion. 
%
%
%
%

\end{document}